\documentclass[seceq]{ptptex}

\usepackage{graphicx}
\title{Excited states in Richardson pairing model: 'probabilistic' approach
}

\author{%       %Use \scshape for the family name.
W. V. \textsc{Pogosov}%
}

\inst{Institute for Theoretical and Applied Electrodynamics,
Russian Academy of Sciences, Izhorskaya 13, 125412 Moscow, Russia
}

\abst{Richardson equations can be mapped on the classical
electrostatic problem in two dimensions. We have recently
suggested a new analytical approach to these equations in the
thermodynamical limit, which is based on the 'probability' of the
system of charges to be in a given configuration at the effective
temperature equal to the interaction constant. In the present
paper, we apply this approach to excited states of the Richardson
pairing model. We focus on the equally-spaced situation and
address arbitrary fillings of the energy layer, where interaction
acts. The 'partition function' for the classical problem on the
plane, which is given by Selberg-type integral, is evaluated
exactly. Three regimes for the energy gap are identified, which
can be treated as the dilute regime of pairs, BCS regime, and
dilute regime of holes. }

\PTPindex{010, 062, 368}  %Input the subject index(es) of your paper,
\begin{document}

\maketitle

\section{Introduction}

Bardeen-Cooper-Schrieffer (BCS) theory plays a very important role
in the microscopic description of
superconductivity.\cite{BCS,Bogoliubov,Schrieffer} \ As shown by
Richardson,\cite{Rich1} \ BCS Hamiltonian turns out to be exactly
solvable. By staying in the canonical ensemble, Richardson managed
to find a many-body wave function, which depends on the set of
energy-like quantities (rapidities) and provides an exact solution
of the Schr\"{o}dinger equation. The number of rapidities is equal
to the number of Cooper pairs in the system, while the Hamiltonian
eigenvalue is given by their sum. Rapidities satisfy the system of
nonlinear algebraic equations, now called Richardson equations.
The resolution of these equations is a formidable task. More
recently, it was shown that Richardson equations can be derived
from the algebraic Bethe-ansatz approach.\cite{Pogosyan} \ They
are also closely related to the well-known Gaudin
model\cite{Gaudin} and Chern-Simons theories.\cite{CFT} \
Richardson equations are now widely used to study numerically
superconducting state in nanometer-scale
systems.\cite{Dukel,Braun} \ In particular, powerful tools of the
quantum inverse scattering method were used to compute correlation
functions\cite{Osterloh} in such systems.

It is quite remarkable that Richardson equations can be mapped
onto the classical electrostatic problem in the
plane.\cite{Gaudin,Rich3} \ Namely, energy-like quantities may be
treated as coordinates of interacting charged particles, which are
placed into the external electric field. Equilibrium positions of
these charges are then equivalent to solutions of Richardson
equations. The origin of this highly remarkable example of
quantum-to-classical correspondence is unclear.

We recently pushed further\cite{Pogosov} the analogy with the
classical electrostatic problem by introducing the occupation
'probability' for the system of charges at the effective
'temperature' given by the interaction amplitude, which goes to
zero in the thermodynamical limit. This leads to the new approach
to treat analytically Richardson equations. Namely, one can
reconstruct an information on the location of the center of masses
for the free charges by using an integration instead of a
straightforward resolving the equations. This is done by
constructing the 'partition function', given by the Selberg-type
integral, so that the energy of the initial \textit{quantum}
problem is determined by the logarithmic derivative of this
\textit{classical} quantity with respect to the inverse
temperature. Note that Selberg integrals are familiar in conformal
field theory and in random-matrix models. This fact suggests an
interesting link with these subjects.

In the present paper, we extend the 'probabilistic' approach to
excited states of Richardson pairing model, which amounts
manipulating more sophisticated and general Selberg-type
integrals. The 'partition function' for the deterministic problem
is found analytically by converting such an integral into a
coupled binomial sum, evaluated using combinatorial properties of
Vandermonde matrix. We then calculate energy difference (gap)
between excited and ground states, which is a more subtle quantity
than the ground state energy itself. At the same time, this
quantity is much more important, since it is not possible to test
ground state energy experimentally in contrast to the gap. We
focus on the equally-spaced model and treat arbitrary fillings of
the energy interval, where attraction between up and down spin
electrons acts. This can be seen\cite{We} as a toy model for the
density-induced crossover between individual fermionic molecules
and a dense regime of BCS pairs.\cite{Leggett,Eagles} \ We
identify three different regimes for the energy of the excited
state. The first one corresponds to low densities of pairs; the
excitation energy in this case is controlled by the single-pair
binding energy. The second regime corresponds to the dense regime
of pairs. It is described by the gap of BCS type, which has a
cooperative origin. The third regime is a 'superdense' regime of
pairs or a dilute regime of holes, which is again controlled by
the single-pair binding energy.

The paper is organized as follows. In \S  2, we briefly formulate
the problem and outline basic ingredients of our method. In \S  3,
we calculate the 'partition function'. In \S  4, we discuss our
results and we conclude in \S  5.

\section{General formulation}

We consider electrons with up and down spins, which interact
through the BCS potential. The Hamiltonian is given by
\begin{equation}
H=\sum_{\mathbf{k}}\varepsilon _{\mathbf{k}}\left(
a_{\mathbf{k}\uparrow
}^{\dagger }a_{\mathbf{k}\uparrow }+a_{\mathbf{k}\downarrow }^{\dagger }a_{%
\mathbf{k}\downarrow }\right)
-V\sum_{\mathbf{k},\mathbf{k}^{\prime }}a_{\mathbf{k}^{\prime
}\uparrow }^{\dagger }a_{-\mathbf{k}^{\prime }\downarrow }^{\dagger }a_{-%
\mathbf{k}\downarrow }a_{\mathbf{k}\uparrow }. \label{Hamil}
\end{equation}%

We stay in the canonical ensemble, i.e., with the number of
electrons fixed. In this case, the eigenvalue of Hamiltonian,
given by equation (\ref{Hamil}), can be represented as a sum of
$N$ energy-like quantities $R_{j}$, which satisfy the system of
$N$ coupled Richardson equations
\begin{equation}
1=\sum_{\mathbf{k}}\frac{V}{2\varepsilon
_{\mathbf{k}}-R_{j}}+\sum_{l,l\neq j}\frac{2V}{R_{j}-R_{l}}.
\label{Richardson}
\end{equation}

It is assumed that the attraction between spin up and spin down
electrons acts only for the free electron states having kinetic
energies confined between $\varepsilon _{F_{0}}$ and $\varepsilon
_{F_{0}}+\Omega $; the same therefore applies to the sums in the
right-hand side (RHS) of equation (\ref{Richardson}). The first
quantity $\varepsilon _{F_{0}}$, within the traditional BCS
framework, can be associated with the Fermi energy of the filled
Fermi sea of noninteracting electrons, while $\Omega /2$ is the
Debye frequency. We then also assume that all the energies of free
electron states are different and they are distributed
equidistantly within this potential layer. Such a situation is
usually refereed to as the equally-spaced model. It can be
understood in terms of a constant density of energy states. The
distance between two nearest levels is given by $1/\rho $, where
$\rho $ is the density of energy states. This means that we
consider a situation with the discrete energy spectrum, but the
distance between nearest energy levels for noninteracting
electrons goes to zero in the thermodynamical limit, so that it
becomes quasi-continuous. Thus, there are in total $ N_{\Omega }=
\rho \Omega $ available energy states of each spin direction in
the potential layer. We fill this potential layer by $N$ pairs. In
the usual BCS configuration, $N=N_{\Omega }/2$ (half-filling),
while we address the arbitrary filling corresponding to
macroscopic $N$ (finite density of pairs). In other words, we
consider a thermodynamical limit, for which the dimensionless
interaction constant $v \equiv \rho V$ stays independent on $N$,
as well as the filling $ N / N_{\Omega }$, whereas $ \rho \sim N$.
Now we take the limit $ N \rightarrow \infty$. Changing the
filling should be considered as a toy model for the
density-induced crossover from the dilute regime of pairs, when
the filling is small, to the dense regime, when the filling is
increased.\cite{Geyer,We} \ This crossover attracts a lot of
attention in the field of ultracold gases. In addition, it can be
relevant for high-$T_{c}$ cuprates.\cite{Levin,Gantmakher} \ Such
a toy model also helps to establish a link between the single-pair
problem solved by Cooper and many-pair BCS condensate. It was
argued\cite{Geyer} that this model can be directly applicable to
some semiconductors. Moreover, it has obvious similarities with
the well-known Eagles model\cite{Eagles}, which however does not
assume constant density of states.

There are two types of lowest excited states in Richardson
approach to the pairing problem. Excitations of the first kind
correspond to one of the rapidities having zero imaginary part and
located between two free electron levels, i.e., in a
quasi-continuum spectrum. Excitations of the second type
correspond to the presence of unpaired electrons, which can appear
for instance due to the breaking of one of the pairs. These
unpaired electrons then block the one-electronic states, which
they populate.\cite{BCS} \ Their role is to bring their own bare
kinetic energy to the total energy of the system, and, moreover,
to modify the energy of the remaining set of pairs. Here we
consider this second type of excited states by addressing a system
of $N$ pairs plus one unpaired electron. At the end, we will show
how, in the thermodynamical limit, the energy of the excited state
of the first kind can be found by reducing the problem to the
excited state of the second kind. Note that, within the
variational approach of Ref. \citen{BCS}, excitations of the first
type were called 'real pairs', while excitations of the second
type correspond to 'broken pairs'. Within the Bogoliubov approach
both types of excitations are handled on the same
footing,\cite{Bogoliubov} \ without making any distinction between
them.

The role of a single unpaired electron is to block the occupied
state: No scattering occurs, since BCS interaction potential
couples only electrons with opposite spins and
momenta.\cite{Schrieffer,Rich1} \ Hence, this state must be
excluded from the sums appearing in Richardson equations
(\ref{Richardson}). We denote the total lowest possible energy of
the system of $N$ pairs plus one unpaired electron with momentum
$\textbf{p}$ as $E_{N,\textbf{p}}$.  For the energy of $N$ pairs
in presence of the unpaired electron we use $E_{R,\textbf{p}}$.
The latter quantity can be found from Richardson equations, where
the corresponding blocked state is excluded from the sums.
Consequently, $E_{N,\textbf{p}}=E_{R,\textbf{p}}+ \varepsilon
_{\mathbf{p}}$. The ground state energy of $N$ pairs is denoted as
$E_{N}$. Within the 'probabilistic' approach, $E_{N}$ was
calculated in Ref. \citen{Pogosov}.

It was shown long time ago that Richardson equations have a
remarkable electrostatic analogy.\cite{Rich3,Gaudin} \ Consider
the function $E_{class}(\left\{ R_{j}\right\} )$, given by
\begin{equation}
E_{class}(\left\{ R_{j}\right\} )=2\left(
\sum_{j}ReR_{j}+V\sum_{j,\mathbf{k}}\ln \left\vert 2\varepsilon
_{\mathbf{k}}-R_{j}\right\vert -2V\sum_{j,l,j<l}\ln \left\vert
R_{l}-R_{j}\right\vert \right),  \label{EnergyCharges}
\end{equation}%
which can be interpreted as the energy of $N$ free classical
particles with electrical charges $2\sqrt{V}$ located on the plane
with coordinates given by (Re $R_{j} $, Im $R_{j}$). Free
particles are placed into a uniform external electric field, which
gives rise to the force $(-2,0)$ acting on each particle. In
addition, free charges repeal each other, and they are also
attracted to $N_{\Omega }$ fixed particles each having a charge
$-\sqrt{V}$ and located at $(2 \varepsilon _{\mathbf{k}},0)$.
Richardson equations can be formally written as the equilibrium
condition for the system of $N$ free charges. This can be seen by
splitting $E_{class}(\left\{ R_{j}\right\} )$ as $W(\left\{
R_{j}\right\} )+W(\left\{ R_{j}^{\ast }\right\} )$ with
\begin{equation}
W(\left\{ R_{j}\right\} )=\sum_{j}R_{j}+V\sum_{j,\mathbf{k}}\ln
\left( 2\varepsilon _{\mathbf{k}}-R_{j}\right)
-2V\sum_{j,l,j<l}\ln \left( R_{l}-R_{j}\right)  \label{W}
\end{equation}%
and by considering conditions $\partial W(\left\{ R_{j}\right\}
)/\partial R_{j}=0$, from which Richardson equations
(\ref{Richardson}) follow. Note that the equilibrium of this
system of free charges is not stable. If we treat the excited
state with one level blocked, the corresponding term must be, of
course, excluded from the sums in the RHS of equations
(\ref{EnergyCharges}) and (\ref{W}), while, in the case of the
ground state, the summation runs over all terms.

We recently suggested an idea to push further the analogy between
the initial quantum problem and the classical problem of Coulomb
plasma in two dimensions by considering occupation 'probabilities'
$S(\left\{ R_{j}\right\} )=\exp \left( -W(\left\{
R_{j}\right\}/T_{eff} \right)$ at the effective 'temperature'
$T_{eff}\equiv V$. The effective 'temperature', in the
thermodynamical limit, goes to zero as $1/N$, so one can
reconstruct an information about the sum of energy-like quantities
in equilibrium without solving Richardson equations directly, but
using an integration of $S(\left\{ R_{j}\right\} )$ over
${R_{j}}$, in a spirit of usual thermodynamics, except of the
facts that integration is performed over half of degrees of
freedom (one-dimensional integration over each $R_{j}$) and
$S(\left\{ R_{j}\right\} )$ is a meromorphic function, which
allows one to deform integration paths. The approach has some
similarities with the large-$N$ expansion for the Dyson
gas.\cite{Zabrodin}

In the case of one level blocked, taking into account an above
expression of $S(\left\{ R_{j}\right\} )$, we write it in the form
\begin{equation}
S(\left\{ R_{j}\right\} )=\frac{\prod_{j,l,j<l}(R_{l}-R_{j})^{2}}{%
\prod_{j=1}^{N}\prod_{\mathbf{k\neq p}}(2\varepsilon _{\mathbf{k}}-R_{j})}%
\exp \left( -\frac{\sum_{j=1}^{N}R_{j}}{V}\right), \label{Sexc}
\end{equation}
where the blocked level $2\varepsilon
_{\mathbf{p}}=2\varepsilon_{F_{0}}+2n_{\mathbf{p}}/\rho$ has been
excluded from the product in the denominator. The 'probability'
can be represented as
\begin{equation}
S(\left\{ R_{j}\right\} )=\frac{\left\{
 \prod_{j,l,j<l}(R_{l}-R_{j})^{2}\right\}
\prod_{j}
(2\varepsilon _{\mathbf{p}}-R_{j})}{%
\prod_{j=1}^{N}\prod_{\mathbf{k}}(2\varepsilon _{\mathbf{k}}-R_{j})}%
\exp \left( -\frac{\sum_{j=1}^{N}R_{j}}{V}\right). \label{Sexc1}
\end{equation}

Note that, while $S(\left\{ R_{j}\right\} )$ for the ground state
has analogies with the square of Laughlin wave function for the
ground state,\cite{Pogosov} \ equation (\ref{Sexc1}) resembles
Laughlin wave function for excited states, since it has an
additional factor of the same type, $\prod_{j} (2\varepsilon
_{\mathbf{p}}-R_{j})$, compared to the ground-state $S$.

Next, we perform a partial-fraction decomposition of $S$ and
rewrite it as
\begin{eqnarray}
S(\left\{ R_{j}=2\varepsilon _{F_{0}}-r_{j}\right\} )=\exp \left( \frac{%
-2N\varepsilon _{F_{0}}}{V}\right)
\left[\prod_{l,j,l>j}(r_{l}-r_{j})^{2}\right]
\notag \\
\times \sum_{n_{1},n_{2}...,n_{N}=0}^{N_{\Omega }}
 \left(
\prod_{j=1}^{N}(-1)^{n_{j}}(r_{j}+\frac{2n_{\mathbf{p}}}{\rho})\binom{N_{\Omega
}}{n_{j}}\frac{\exp (r_{j}/V)}{r_{j}+\frac{2n_{j}}{\rho }}\right),
\label{Sdecomp}
\end{eqnarray}%
where $\binom{N_{\Omega }}{n_{j}}$ is the binomial coefficient.
Here and below we drop all irrelevant prefactors independent on
$V$.

To apply our technique, we must be sure that $S(\left\{
R_{j}\right\} )$ goes to zero, as an imaginary part of any of the
$R$ tends to infinity. This criterium is satisfied provided that
the filling is not larger than $1/2$, as follows from equation
(\ref{Sexc1}). To address larger fillings, we should switch to the
representation in terms of holes, as it has been done in Ref.
\citen{Pogosov}.

\section{Evaluation of 'partition function'}

\subsection{'Partition function' as the binomial sum}

We now introduce a 'partition function', which is given by the
Selberg-type multidimensional integral
\begin{equation}
Z= \int S(\left\{ R_{j}\right\} ) dR_{j}, \label{partit}
\end{equation}
where an integration is performed over the whole set $\left\{
R_{j}\right\}$ along the contour for each $R_{j}$, shown in Fig.
1. The integration path avoids all the poles of $S$ (corresponding
to $2\varepsilon _{\mathbf{k}}$) and then reconnects via the
semicircle of infinite radius, along which $S(\left\{
R_{j}\right\} )$ is zero. The sum $E_{R,\textbf{p}}$ of
energy-like quantities $R$ in the equilibrium, which is a quantity
of interest, is expressed through the logarithmic derivative of
$Z$ with respect to the inverse 'temperature' $V$
\begin{equation}
E_{R,\textbf{p}}=-\frac{\partial }{\partial \left(
\frac{1}{V}\right) }\ln Z. \label{logderiv}
\end{equation}%

\begin{figure}
\centerline{\includegraphics[width=5 cm] {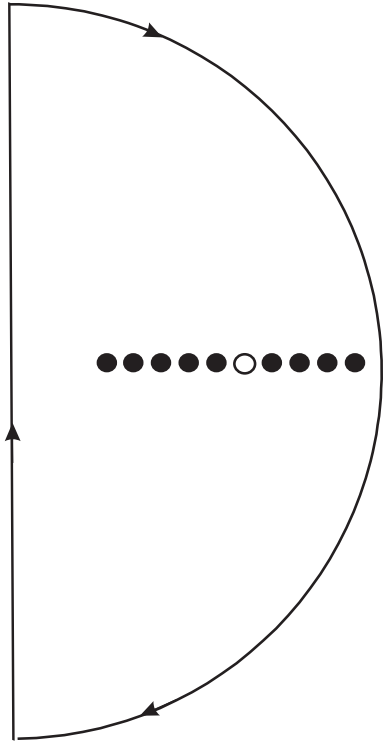}}
\caption{The schematic plot of complex values of each $R_{j}$.
Filled circles show locations of energy levels for free electrons.
Open circle indicates a missing (blocked) level. Solid line
corresponds to the integration path for each $R_{j}$. }
\label{fig:1}
\end{figure}

At this stage, we substitute equation (\ref{Sdecomp}) to equation
(\ref{partit}) and integrate using residues. We easily get
\begin{equation}
Z=\exp (-2N\varepsilon _{F_{0}}/V)z, \label{Zexccap}
\end{equation}%
where
\begin{equation}
z=\sum_{n_{1},n_{2}...,n_{N}=0}^{N_{\Omega}}
\left( \prod_{j=1}^{N}(-1)^{n_{j}}\sigma ^{n_{j}}\binom{N_{\Omega }}{n_{j}}%
(n_{j}-n_{\mathbf{p}})\right) \left(
\prod_{l,j,l>j}(n_{l}-n_{j})^{2} \right), \label{Zexcsmall}
\end{equation}
where $\sigma = \exp (-2/v)$.

\subsection{Vandermonde determinants}

It is known that $\prod_{l,j,l>j}(n_{l}-n_{j})$ entering equation
(\ref{Zexcsmall}) can be represented as the determinant of the
Vandermonde matrix
\begin{equation}
\prod_{l,j,l>j}(n_{l}-n_{j})\equiv V_{0}(\left\{ n_{j}\right\}
)=\det
\begin{bmatrix}
1 & 1 & 1 & \ldots & 1 \\
n_{1} & n_{2} & n_{3} & \ldots & n_{N} \\
n_{1}^{2} & n_{2}^{2} & n_{3}^{2} & \ldots & n_{N}^{2} \\
\ldots & \ldots & \ldots & \ldots & \ldots \\
n_{1}^{N-1} & n_{2}^{N-1} & n_{3}^{N-1} & \ldots & n_{N}^{N-1}%
\end{bmatrix}%
.  \label{vander}
\end{equation}
Consequently, we can rewrite the total product
$\prod_{l,j,l>j}(n_{l}-n_{j})^{2}$ as\\
$(-1)^{N(N-1)/2}V_{0}(\left\{ {n_{j}}\right\} )V_{0}(\left\{
N_{\Omega }-n_{j}\right\} )$.

Instead of the standard monomials, we represent $V_{0}(\left\{
{n_{j}}\right\} )$ in terms of Pochhammer symbols
$(n)_{a}=n(n-1)...(n-a+1)$, while $(n)_{0}\equiv1$. This can be
done by iterative substraction of rows of Vandermonde matrix,
which does not change the determinant and ultimately leads to
\begin{equation}
V_{0}(\left\{ {n_{j}}\right\} )=\det
\begin{bmatrix}
(n_{1})_{0} & (n_{2})_{0} & (n_{3})_{0} & \ldots & (n_{N})_{0} \\
(n_{1})_{1} & (n_{2})_{1} & (n_{3})_{1} & \ldots & (n_{N})_{1} \\
(n_{1})_{2} & (n_{2})_{2} & (n_{3})_{2} & \ldots & (n_{N})_{2} \\
\ldots & \ldots & \ldots & \ldots & \ldots \\
(n_{1})_{N-1} & (n_{2})_{N-1} & (n_{3})_{N-1} & \ldots & (n_{N})_{N-1}%
\end{bmatrix}%
,  \label{vander1}
\end{equation}%
while $V_{0}(\left\{ N_{\Omega }-n_{j}\right\} )$ can be
represented in a similar form with $n_{j}$ changed into $N_{\Omega
}-n_{j}$.

Next, we insert $(n_{j}-n_{\mathbf{p}})$, which appears in
equation (\ref{Zexcsmall}), into the Vandermonde matrix in the
following manner
\begin{eqnarray}
\left( \prod_{j=1}^{N}(n_{j}-n_{\mathbf{p}}) \right) \left(
\prod_{j,l,l>j}(n_{l}-n_{j}) \right) \notag \\
=\det
\begin{bmatrix}
(n_{1})_{0}(n_{1}-n_{\mathbf{p}}) &
(n_{2})_{0}(n_{2}-n_{\mathbf{p}}) & \ldots &
(n_{N})_{0}(n_{N}-n_{\mathbf{p}})
\\
(n_{1})_{1}(n_{1}-n_{\mathbf{p}}) &
(n_{2})_{1}(n_{2}-n_{\mathbf{p}}) & \ldots &
(n_{N})_{1}(n_{N}-n_{\mathbf{p}})
\\
\ldots & \ldots & \ldots & \ldots \\
(n_{1})_{N-1}(n_{1}-n_{\mathbf{p}}) & (n_{2})_{N-1}(n_{2}-n_{\mathbf{p}}) & \ldots & (n_{N})_{N-1}(n_{N}-n_{\mathbf{p}})%
\end{bmatrix}
\label{Vanderexc1}
\end{eqnarray}%
Let us now extract $n_{\mathbf{p}}$ from the first row of the
above matrix as
\begin{eqnarray}
&\det&
\begin{bmatrix}
(n_{1})_{1} & (n_{2})_{1} & \ldots & (n_{N})_{1} \\
(n_{1})_{1}(n_{1}-n_{\mathbf{p}}) &
(n_{2})_{1}(n_{2}-n_{\mathbf{p}}) & \ldots & (n_{N})_{1}(n_{N}-p)
\\
\ldots & \ldots & \ldots & \ldots \\
(n_{1})_{N-1}(n_{1}-n_{\mathbf{p}}) & (n_{2})_{N-1}(n_{2}-n_{\mathbf{p}}) & \ldots & (n_{N})_{N-1}(n_{N}-n_{\mathbf{p}})%
\end{bmatrix}
\notag \\
-n_{\mathbf{p}}&\det&
\begin{bmatrix}
(n_{1})_{0} & (n_{2})_{0} & \ldots & (n_{3})_{0} \\
(n_{1})_{1}(n_{1}-n_{\mathbf{p}}) &
(n_{2})_{1}(n_{2}-n_{\mathbf{p}}) & \ldots &
(n_{N})_{1}(n_{N}-n_{\mathbf{p}})
\\
\ldots & \ldots & \ldots & \ldots \\
(n_{1})_{N-1}(n_{1}-n_{\mathbf{p}}) & (n_{2})_{N-1}(n_{2}-n_{\mathbf{p}}) & \ldots & (n_{N})_{N-1}(n_{N}-n_{\mathbf{p}})%
\end{bmatrix}
\notag \\
&\equiv& D_{1}-n_{\mathbf{p}}D_{2}. \label{Vanderexc2}
\end{eqnarray}%
The first term, $D_{1}$, can be rewritten as
\begin{eqnarray}
&&D_{1}=\det
\begin{bmatrix}
(n_{1})_{1} & (n_{2})_{1} & \ldots & (n_{3})_{1} \\
(n_{1})_{2} & (n_{2})_{2} & \ldots & (n_{N})_{2} \\
\ldots & \ldots & \ldots & \ldots \\
(n_{1})_{N-1}(n_{1}-n_{\mathbf{p}}) & (n_{2})_{N-1}(n_{2}-n_{\mathbf{p}}) & \ldots & (n_{N})_{N-1}(n_{N}-n_{\mathbf{p}})%
\end{bmatrix}
\\
&&-(n_{\mathbf{p}}-1)\det
\begin{bmatrix}
(n_{1})_{1} & (n_{2})_{1} & \ldots & (n_{N})_{1} \\
(n_{1})_{1} & (n_{2})_{1} & \ldots & (n_{N})_{1} \\
\ldots & \ldots & \ldots & \ldots \\
(n_{1})_{N-1}(n_{1}-n_{\mathbf{p}}) &
(n_{2})_{N-1}(n_{2}-n_{\mathbf{p}}) & \ldots &
(n_{N})_{N-1}(n_{N}-n_{\mathbf{p}})
\notag \\
\end{bmatrix}%
\label{Vanderexc3}
\end{eqnarray}%
where we have extracted $n_{\mathbf{p}}-1$ from the second row of
the initial matrix. We immediately see that the last determinant
is exactly zero, since two rows of its matrix are the same. We
then extract $n_{\mathbf{p}}-2$ from the third row of the
remaining matrix in the RHS of equation (\ref{Vanderexc3}) and
repeat our arguments. We follow a similar iterative procedure for
$D_{2}$. It is rather straightforward to see that only terms
proportional to $(n_{\mathbf{p}})_{m}$ do survive at the end,
since all the other terms are proportional to determinants of
matrices with repeating rows. We finally arrive to the following
identity
\begin{equation}
\left( \prod_{j=1}^{N}(n_{j}-n_{\mathbf{p}}) \right)
\left( \prod_{j,l,l>j}(n_{l}-n_{j}) \right)=%
\sum_{m=0}^{N}(-1)^{m}(n_{\mathbf{p}})_{m} V_{N-m}(\left\{
{n_{j}}\right\} ) \label{Coulombplus}
\end{equation}%
where $V_{m}(\left\{ {n_{j}}\right\} )$ are determinants of the
matrices, which are obtained from the matrix of the RHS of
equation (\ref{vander1}) by increasing indices of Pochhammer
symbols in the last $m$ rows by one, as $(n_{j})_{a}\rightarrow
(n_{j})_{a+1}$.

\subsection{Auxiliary identities}

To go further, we provide some auxiliary identities, which are
going to greatly facilitate calculations. The starting identity is
\begin{equation}
z_{a,b}\equiv \sum_{n=0}^{N_{\Omega }}(-1)^{n}\sigma ^{n}\binom{N_{\Omega }}{%
n}(n)_{a}(N_{\Omega }-n)_{b}=\sigma ^{a}(1-\sigma )^{N_{\Omega }-a-b}%
(-1)^{a} \frac{N_{\Omega }!}{(N_{\Omega }-a-b)!},  \label{eat}
\end{equation}%
where $a+b\leq N_{\Omega }$. It can be obtained by observation
that first $a$ terms, as well as last $b$ terms of the initial sum
are zero, and then by replacing Pochhammer symbols by ratios of
factorials.

Let us now focus on a product of sums for each $%
n_{j}$, every sum being similar to the one of equation
(\ref{eat}). We can write such a product as
\begin{eqnarray}
z_{a_{1},b_{1}}...z_{a_{N},b_{N}} &=&\sigma
^{\sum_{j=1}^{N}a_{j}}(1-\sigma )^{NN_{\Omega
}-\sum_{j=1}^{N}(a_{j}+b_{j})}(-1)^{\sum_{j=1}^{N}a_{j}}
\notag \\
\times \prod_{j=1}^{N}\frac{%
N_{\Omega }!}{(N_{\Omega }-a_{j}-b_{j})!}.  \label{eating}
\end{eqnarray}%
The dependence of this quantity on $V$ (via $\sigma$) is through
two first factors.
Their dependencies on sets ${a_{j}}$ and ${b_{j}}$ is only by $%
\sum_{j=1}^{N}a_{j}$ and $\sum_{j=1}^{N}b_{j}$, which actually are
degrees of polynomials $\prod_{j=1}^{N}(n_{j})_{a_{j}}$ and
$\prod_{j=1}^{N}(N_{\Omega }-n_{j})_{b_{j}}$, respectively.

\subsection{'Partition function': hypergeometric series}

Let us substitute equation (\ref{Coulombplus}) back to equation
(\ref{Zexcsmall}). To proceed in calculations, we have to consider
products of the form $V_{m}(\left\{ {n_{j}}\right\} )V_{0}(\left\{
N_{\Omega }-n_{j}\right\} )$. It is easy to see from equation
(\ref{vander1}) that each of them (for a given $m$) can be
represented as a linear combination of polynomials of the form
$\left( \prod_{j=1}^{N}(n_{j})_{a_{j}}\right) \left(
\prod_{j=1}^{N}(N_{\Omega }-n_{j})_{b_{j}}\right) $ with the unique $%
\sum_{j=1}^{N}a_{j} = m+N(N-1)/2$ and
$\sum_{j=1}^{N}b_{j}=N(N-1)/2$ for each polynomial. These two
numbers just give the degrees of the polynomials $V_{m}(\left\{
{n_{j}}\right\} )$ and $V_{0}(\left\{ N_{\Omega }-n_{j}\right\}
)$, respectively.
 According to equation (\ref{eating}), after substitution of these products to equation
(\ref{Zexcsmall}) and performing summations, we get the same
dependence of the result on $\sigma$ for any of these polynomials
(at a given $m$). Hence, we can write
\begin{equation}
z=\sigma ^{N(N+1)/2}\left( 1-\sigma \right) ^{N(N_{\Omega
}-N)}\sum_{m=0}^{N}(n_{\mathbf{p}})_{m}\left( \sigma ^{-1}-1
\right) ^{m}\alpha _{m}, \label{zalfa}
\end{equation}%
where $\alpha _{m}$ are unknown numbers independent on $\sigma $,
which have a combinatorial origin.

We avoid a direct calculation of ${\alpha _{m}}$ by using a trick,
which is based on the well-known rule: $\binom{N_{\Omega
}}{n}=\binom{N_{\Omega }}{N_{\Omega }-n}$. Namely, we change
summation variables in equation (\ref{Zexcsmall}) as
$n_{j}^{^{\prime }}=N_{\Omega }-n_{j}$. It is then readily seen
that
\begin{equation}
z(n_{\mathbf{p}},\sigma )=(-1)^{N+NN_{\Omega }}\sigma ^{NN_{\Omega
}}z(N_{\Omega }-n_{\mathbf{p}},\sigma ^{-1}). \label{zswitch}
\end{equation}%
For $z(N_{\Omega }-n_{\mathbf{p}},\sigma ^{-1})$ we can use
equation (\ref{zalfa}) with $n_{\mathbf{p}}\rightarrow N_{\Omega
}-n_{\mathbf{p}}$, $\sigma \rightarrow \sigma ^{-1}$. By doing
this, we arrive to another expression of $z(n_{\mathbf{p}},\sigma
)$ in terms of $(N_{\Omega }-n_{\mathbf{p}})_{m}$,
which is nevertheless equivalent to equation (\ref{zalfa}):%
\begin{equation}
z=\sigma ^{N(N-1)/2}\left( 1-\sigma \right) ^{N(N_{\Omega
}-N)}\sum_{m=0}^{N}(N_{\Omega }-n_{\mathbf{p}})_{m}\left( \sigma
^{-1}-1\right) ^{m}\alpha _{m}. \label{zalfanew}
\end{equation}

The idea is to express Pochhammer symbols of $N_{\Omega
}-n_{\mathbf{p}}$\ in terms of Pochhammer symbols of
$n_{\mathbf{p}}$, then to substitute them to equation
(\ref{zalfanew}) and to compare
the result with equation (\ref{zalfa}). By equating coefficients of Pochhammer symbols of $n_{\mathbf{p}}$%
, we are going to obtain a system of linear equations for $\left\{
\alpha _{m}\right\}
$. We make use of the following relation%
\begin{equation}
(N_{\Omega }-n_{\mathbf{p}})_{m}=\sum_{l=0}^{m}(-1)^{l}(n_{\mathbf{p}})_{l}\binom{m}{l}\frac{%
(N_{\Omega }-l)!}{(N_{\Omega }-m)!}, \label{Pochswitch}
\end{equation}%
which can be trivially checked for $m=0,$ $1$ and then proved by
induction. Inserting it to equation (\ref{zalfanew}) and solving
the system of equations for $\left\{
\alpha _{m}\right\} $, we get%
\begin{equation}
\alpha _{m}=\alpha \binom{N}{m}\frac{1}{(N_{\Omega })_{m}},
\label{alfamalfa}
\end{equation}%
where $\alpha $ is an irrelevant constant, independent on
$\sigma$.

Finally, we obtain $z$ as%
\begin{eqnarray}
z &=&\alpha \sigma ^{N(N+1)/2}\left( 1-\sigma \right)
^{N(N_{\Omega }-N)}z^{\prime},
 \label{zprime}
\end{eqnarray}%
where
\begin{eqnarray}
z^{\prime}=\sum_{m=0}^{N}\binom{N}{m}\frac{(n_{\mathbf{p}})_{m}}{(N_{\Omega })_{m}}%
\left( \sigma ^{-1}-1\right) ^{m}.
 \label{Jacobi}
\end{eqnarray}%
The last expression can be considered as a hypergeometric series.

\subsection{Hypergeometric series: saddle-point method}

Actually, equation (\ref{zprime}) already allows us to find
$E_{R,\textbf{p}}$ in terms of the hypergeometric series. The
resulting expression, however, is essentially untractable from the
perspective of a further analysis. Let us, therefore, try to
transform $z^{\prime}$ into a simpler form.

We note that, for natural numbers $n_{\mathbf{p}}$ and $m$,
$(n_{\mathbf{p}})_{m}=0$ for any $m$, which is larger than
$n_{\mathbf{p}}$. Hence, we can change the upper limit of
summation in equation (\ref{Jacobi}) to $\min(n_{\mathbf{p}},N)$.
After that, all the terms in the sum are nonzero, so that we can
replace Pochhammer symbols of $n_{\mathbf{p}}$ by ratios of
factorials, as
$(n_{\mathbf{p}})_{m}=n_{\mathbf{p}}!/(n_{\mathbf{p}}-m)!$.

We also see that $\sigma^{-1}-1\equiv\exp(2/v)-1$ is always
positive. The last circumstance is very important: it means that
terms in the sum are not oscillating in sign. Therefore, we can
switch from summation to integration. We also utilize asymptotic
expansion for factorials entering both Pochhammer symbols and the
binomial coefficient, since we are interested in macroscopic
numbers, $\sim N$. Note that the important case of
$n_{\mathbf{p}}=0$ must be considered as $n_{\mathbf{p}}=0 \times
N$, although this particular situation can be analyzed separately
without switching to integration, since the derivation turns out
to be quite simple (the results of both approaches are finally the
same). After straightforward algebra, we arrive to the identity
\begin{eqnarray}
z^{\prime}\simeq\int_{0}^{\min(n_{\mathbf{p}},N)} e^{H(m)}dm,
 \label{zprimeint}
\end{eqnarray}%
where
\begin{eqnarray}
H(m)=(N_{\Omega}-m)\ln(N_{\Omega}-m)-m\ln(m)-(N-m)\ln(N-m)
\notag \\
-(n_{\mathbf{p}}-m)\ln(n_{\mathbf{p}}-m)+m\ln \left( \sigma
^{-1}-1\right).
 \label{Hm}
\end{eqnarray}%
To avoid possible confusion, we note that in equation
(\ref{zprimeint}) we have dropped, as usual, an irrelevant
$V$-independent prefactor.

In order to evaluate the integral in the RHS of equation
(\ref{zprimeint}), we use a saddle-point method. It is
straightforward to prove that $H(m)$ has a single maximum within
the integration range, which is attained at
\begin{eqnarray}
m_{0}=\frac{1}{2}\
\left\{(n_{\mathbf{p}}+N)(1-\sigma)+N_{\Omega}\sigma\ \right\}
 \notag \\
 \times \left\{
1-\sqrt{1-\frac{4n_{\mathbf{p}}N(1-\sigma)}{((n_{\mathbf{p}}+N)(1-\sigma)+N_{\Omega}\sigma)^{2}}}\right\}.
 \label{m0}
\end{eqnarray}%
We can also ensure that the position of this maximum is far enough
from both integration limits, since $H^{\prime\prime}(m_{0})\sim
N^{-1}$, while $1/\sqrt{H^{\prime\prime}(m_{0})} \sim \sqrt{N}$
determines the width of the neighborhood of $m_{0}$, which gives
the dominant contribution to the integral. At the same time, both
$m_{0}$ and $(\min(n_{\mathbf{p}},N)-m_{0})$ do scale as $N$.
Together with the fact that $H^{(n)}(m_{0})\sim N^{-n+1}$ at
$n>2$, this enables us to reduce the problem to the simple
Gaussian integration. Keeping leading order in $N$, we obtain
\begin{eqnarray}
\ln z^{\prime}=H(m_{0}).
 \label{logzpr}
\end{eqnarray}%

By finding a logarithmic derivative of $z$, given by equation
(\ref{zprime}), and by adding the kinetic energy of the unpaired
electron $\varepsilon_{F0}+n_{\textbf{p}}/\rho$, we arrive to the
expression of the total energy $E_{N,\textbf{p}}$ of the system of
$N$ pairs and one unpaired electron. After some simple algebra and
using the known expression for the ground state
energy\cite{Pogosov} $E_{N}$, we can present $E_{N,\textbf{p}}$ as
\begin{eqnarray}
E_{N,\textbf{p}}=E_{N}+\mu+\sqrt{\Delta^{2}+(\varepsilon_{\mathbf{p}}-\mu)^{2}},
 \label{final}
\end{eqnarray}%
while
\begin{eqnarray}
\Delta=\frac{2}{\rho}\frac{\sqrt{\sigma}}{1-\sigma}\sqrt{N(N_{\Omega}-N)},
 \label{delta}
\end{eqnarray}%
\begin{eqnarray}
\mu=\varepsilon_{F0}+\frac{N}{\rho}\frac{1+\sigma}{1-\sigma}-\Omega\frac{\sigma}{1-\sigma}.
 \label{mu}
\end{eqnarray}%
Physical meanings of $\Delta$ and $\mu$ will be fixed below.

The formalism presented above is restricted to $N \leq
N_{\Omega}/2$. In order to address configurations with $ N
>N _{\Omega}/2 $, we switch to holes. It is then
straightforward to ensure that equation (\ref{final}) still holds
in this case.

\section{Discussion}

We now consider $E_{N,\textbf{p}}$ as a function of
$\varepsilon_{p}$. In particular, it is of interest to determine
the lowest possible $E_{N,\textbf{p}}$.

First of all, we see that $\mu$, given by equation (\ref{mu}), can
be actually considered as a chemical potential. Indeed, by
calculating $(E_{N}-E_{N-1})/2$, we find that this quantity does
coincide with $\mu$.

The minimum value of $E_{N,\textbf{p}}$ is attained at
$n_{\mathbf{p}}$, which gives a minimum of
$(\varepsilon_{\textbf{p}}-\mu)^{2}$, provided that
$n_{\mathbf{p}}$ is confined between 0 and $N_{\Omega}$. If
$\mu-\varepsilon_{F0}$ also falls into this range, one can always
choose $n_{\mathbf{p}}/\rho=\mu-\varepsilon_{F0}$, so that
$(\varepsilon_{\textbf{p}}-\mu)^{2}$ is zero, while the square
root in the RHS of equation (\ref{final}) reduces to $\Delta$. It
is easy to see that for the half-filling configuration,
$N_{\Omega}=N/2$, the expression of $\Delta$, as given by equation
(\ref{delta}), reproduces precisely the BCS formula for the gap.
This conclusion is also in agreement with the result of the
traditional method to solve Richardson equations in the
large-sample limit, which uses an assumption that energy-like
quantities are arranged into arcs on the complex
plane.\cite{Rich3,Roman,Altshuler} \ The assumption on arcs is
actually deduced from numerical solutions of Richardson equations,
so that, in general case, it is not fully controllable, from our
point of view. Similar approach is also widely used in a broader
context for the solution of the Bethe-ansatz
equations.\cite{Kazakov} \ Another method,\cite{We} \ which is
based on Taylor expansions of sums appearing in Richardson
equations around the known single-pair solution, up to now has
been successfully applied to the ground state only, while its
application to excited states leads to heavy mathematical
problems.

If the chemical potential goes below the lower cutoff
$\varepsilon_{F0}$, i.e., $\mu-\varepsilon_{F0}$ becomes negative,
a constrained minimum of $E_{N,\textbf{p}}$ corresponds to
$n_{\mathbf{p}}=0$. By solving the equation
$\mu-\varepsilon_{F0}=0$, we see that the transition between the
two regimes occurs at $N=N_{0}$, where
\begin{eqnarray}
N_{0}=N_{\Omega}\frac{\sigma}{1+\sigma}.
 \label{N1}
\end{eqnarray}%

If the chemical potential goes above the upper limit of the
potential layer, the minimum energy corresponds to
$\varepsilon_{p}$ located at this upper limit. By solving the
equation $\mu-\varepsilon_{F0}=\Omega$, we find that the
transition to this regime happens at $N_{\Omega}-N_{0}$. This
result is in agreement with the electron-hole symmetry.

Thus, for the minimum of $E_{N,\textbf{p}}$, we have three regimes
\begin{eqnarray}
\min (E_{N,\textbf{p}})=E_{N}+\mu+ \left( \frac{N}{\rho}+\Omega
\frac{\sigma}{1-\sigma} \right),
 \label{Emin1}
\end{eqnarray}%
at $N\leq N_{0}$;
\begin{eqnarray}
\min (E_{N,\textbf{p}})=E_{N}+\mu+\Delta,
 \label{Emin2}
\end{eqnarray}%
at $N_{0} < N < N_{\Omega}-N_{0}$;
\begin{eqnarray}
\min (E_{N,\textbf{p}})=E_{N}+\mu+\left\{
\left(\Omega-\frac{N}{\rho} \right) +\Omega
\frac{\sigma}{1-\sigma} \right\},
 \label{Emin3}
\end{eqnarray}
at $ N_{\Omega}-N_{0} \leq N$.

In the weak-coupling limit, $\sigma\ll 1$, we have: $N_{0} \simeq
N_{\Omega} \sigma$. This value actually corresponds to the density
of pairs, at which their wave functions start to overlap. Such an
overlap can be considered as a signature of the transition from
the isolated-pair regime to the dense condensate.

In the dilute regime, $N < N_{0}$, the excitation energy is
controlled by $\Omega \sigma /(1-\sigma)$, which is nothing but
half the binding energy $\epsilon _{c}$ of an isolated
pair.\cite{We} \ In the dense regime, $N_{\Omega}-N_{0} \geq N>
N_{0}$, it is governed by the BCS gap $\Delta$, which has to be
considered as a collective many-body response of the system to the
appearance of one blocked level. It is of interest to note that
pair binding energy and the gap have similar, but different
dependencies on interaction constant $v$. Namely, in the
weak-coupling limit, the first quantity is proportional to
$exp(-2/v)$, while the second one behaves as $\sim exp(-1/v)$ due
to equation (\ref{delta}). This equation also shows that $\Delta$
is symmetric with respect to the mutual replacement of electron
and holes, so that the electron-hole symmetry again shows up. The
third regime can be considered as a 'superdense' regime of Cooper
pairs, made of electrons, or the dilute regime of Cooper pairs,
made of holes. In this regime, we again see a single-pair binding
energy appearing in $\min (E_{N,\textbf{p}})$. It can be now
understood as a binding energy of a pair made out of holes.

Thus, we have identified three regimes for the energy difference
between $E_{N,\textbf{p}}$ and $E_{N}$ depending on the energy
layer filling. These are a dilute regime of pairs, BCS regime, and
a dilute regime of holes. With changing layer filling, transitions
between these regimes occur smoothly. We have found that only
higher-order derivatives of $E_{N,\textbf{p}}$ with respect to $N$
experience discontinuities upon the transitions.

Up to now, we considered states with only one unpaired electron.
Let us address the energy of the state with two such electrons.
This should allow us to make a comparison with the energy of the
system with all the electrons paired, the total number of
particles being conserved. In principle, in order to find the
energy, we should perform similar calculations, but with two
states blocked. However, we may use a simple trick allowing one to
avoid making such computations. It is rather obvious that it is
energetically favorable for the two unpaired electrons to occupy
two neighboring states rather than to be separated. We then come
back to the electrostatic picture and perform a coarse-graining of
the initial configuration. Namely, we merge couples of neighboring
fixed particles, as well as couples of neighboring free particles
into 'superparticles' with charges being twice larger than initial
ones. We must also increase by a factor of two a homogeneous
forces acting on free charges. This procedure does not change
dominant (extensive) contribution of the total energy. Within this
procedure, two blocked states are converted into a single one, so
that we can map this picture to the above situation. By doing
this, after some straightforward calculations, we finally reach an
expectable result: the difference of energies of the states with
$N$ pairs and $N-1$ pairs plus two unpaired electrons is twice the
square root entering RHS of (\ref{final}).

Finally, let us discuss another type of excited states, which
corresponds to one of the energy-like quantities trapped between
two one-electronic levels. The energy of such a state can be again
handled using electrostatic analogy without making detailed
calculations. Namely, we note that the role of the free charge
trapped is to compensate \textit{two} fixed charges, between which
it is accommodated, since each of them has an opposite charge, but
twice smaller in absolute value. We again map this configuration
to the previous one with two states blocked and find the same gap,
within dominant terms in $1/N$.

\section{Conclusions}

Richardson equations provide an exact solution for the BCS pairing
Hamiltonian. These equations are deterministic and posses a
well-known electrostatic analogy. Therefore, one can convert the
problem of their resolution to the probabilistic problem, as it
was recently suggested in Ref. \citen{Pogosov} for the ground
state of the initial quantum problem. This approach avoids
assumption that Richardson solutions in the large-$N$ limit are
arranged in arcs on the complex plane.

In the present paper, we applied this treatment to excited states
with the focus on the equally-spaced model and the thermodynamical
limit. We have considered arbitrary fillings of the energy
interval, where the attractive potential acts. The 'partition
function' for the deterministic problem has been found
analytically by converting the Selberg-type integral into coupled
binomial sum, evaluated using combinatorial properties of
Vandermonde matrix.

For the energy difference between the first excited state and the
ground state (energy gap), three regimes have been identified,
which can be considered as the dilute regime of pairs, BCS regime,
and dilute regime of holes. Explicit expressions have been
derived. Transitions between these regimes occur smoothly,
accompanied by only weak singularities. The results supports the
BCS result for the half-filling.

\section*{Acknowledgements}
The author acknowledges numerous discussions with Monique
Combescot. This work was supported by RFBR (project no.
12-02-00339), joint Russian-French programme (RFBR-CNRS project
no. 12-02-91055), Dynasty Foundation, and, in parts, by the French
Ministry of Education.

\end{document}